# Physical Model for the Evolution of the Genetic Code

**Tatsuro Yamashita・Osamu Narikiyo**
*Department of Physics, Kyushu University, Fukuoka 810-8560, Japan*

**Abstract**
We propose a physical model to describe the mechanisms of two major scenarios of the genetic code evolution, the codon capture and ambiguous intermediate scenarios, in a consistent manner. We sketch the lowest dimensional version of our model, a minimal model, by introducing a physical quantity, codon level. On the basis of the hierarchical structure of the codon levels two scenarios are classified into two different routes of the evolutional process. In the case of the ambiguous intermediate scenario we perform a simulation implemented cost selection of amino acids and confirm a rapid transition of the code change. Such rapidness reduces uncomfortableness of the non-unique translation of the code at intermediate state that is the weakness of the scenario.

## 1 Introduction

The genetic code is the central element of every biological phenomenon. It was supposed to be frozen at first, but it turned out to be evolvable after the discovery of nonstandard code in mitochondria [1]. After that several scenarios for the evolution of the genetic code [1-3], namely the reassignment of codons to amino acids, have been proposed. Among them the codon capture [1,4] and ambiguous intermediate [5,6] scenarios are major ones.

In the text-book argument [1] the possibility of the latter scenario was questioned. However, recent analysis [7] for observed codon reassignments supports both scenarios according to the situation. The codon capture scenario is supported by observed codon reassignments from stop to an amino acid. On the other hand, the ambiguous intermediate scenario is supported by observed codon reassignments from one amino acid to another.

Thus in this paper by introducing a concept of codon level we try to construct a physical model which realizes both scenarios in a consistent manner. On the basis of our physical model we can discuss the mechanism of the genetic code evolution. As far as we know, a study of the mechanism by physical arguments has not been reported.

This paper is organized as follows. In Section 2 we summarize two scenarios and introduce our physical model. Some simulation results on the basis of our model are discussed in Section 3 focusing on the time dependence of the genetic code evolution. Our conclusion is given in Section 4.

## 2 Model

### 2-1 Codon Capture and Ambiguous Intermediate Scenarios

The abstract of the codon capture [1,4] and ambiguous intermediate [5,6] scenarios are shown in Fig. 1. The codon reassignments mainly arise from alterations in tRNA. According to the position of the tRNA alteration, favored scenario changes as shown in Fig. 2. The situation is summarized [2,3] as follows and we adopt this summary as a basis of our model. If the alteration occurs at the anticodon in tRNA, the codon reassignment proceeds according as the codon capture scenario. On the other hand, if the alteration occurs at locations other than the anticodon in tRNA, the codon reassignment proceeds according to the ambiguous intermediate scenario.

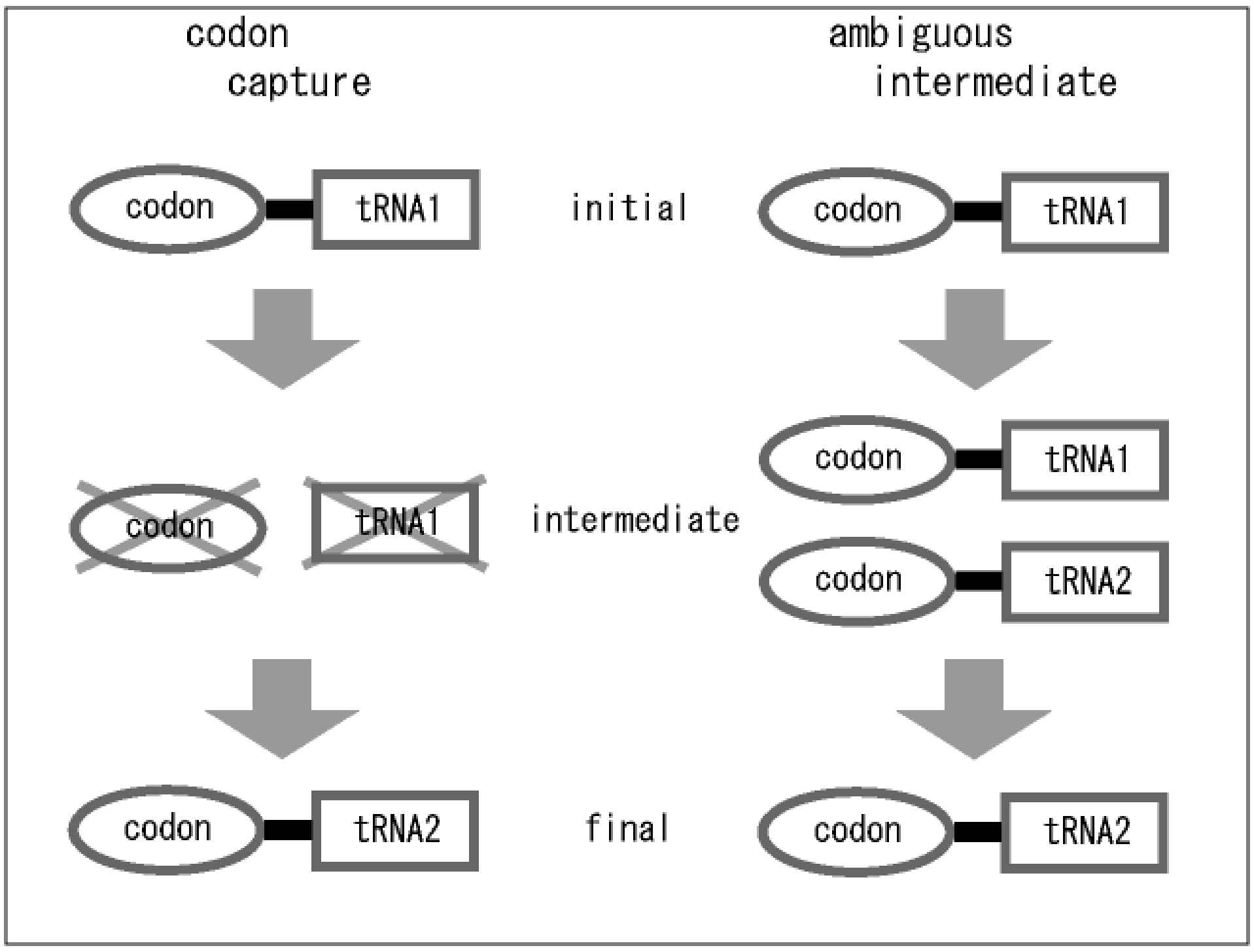

**Fig. 1** Two major scenarios of genetic code evolution: codon capture and ambiguous intermediate scenarios. In the intermediate state both codon and tRNA disappear in codon capture scenario but a codon is recognized by two different tRNAs in ambiguous intermediate scenario.

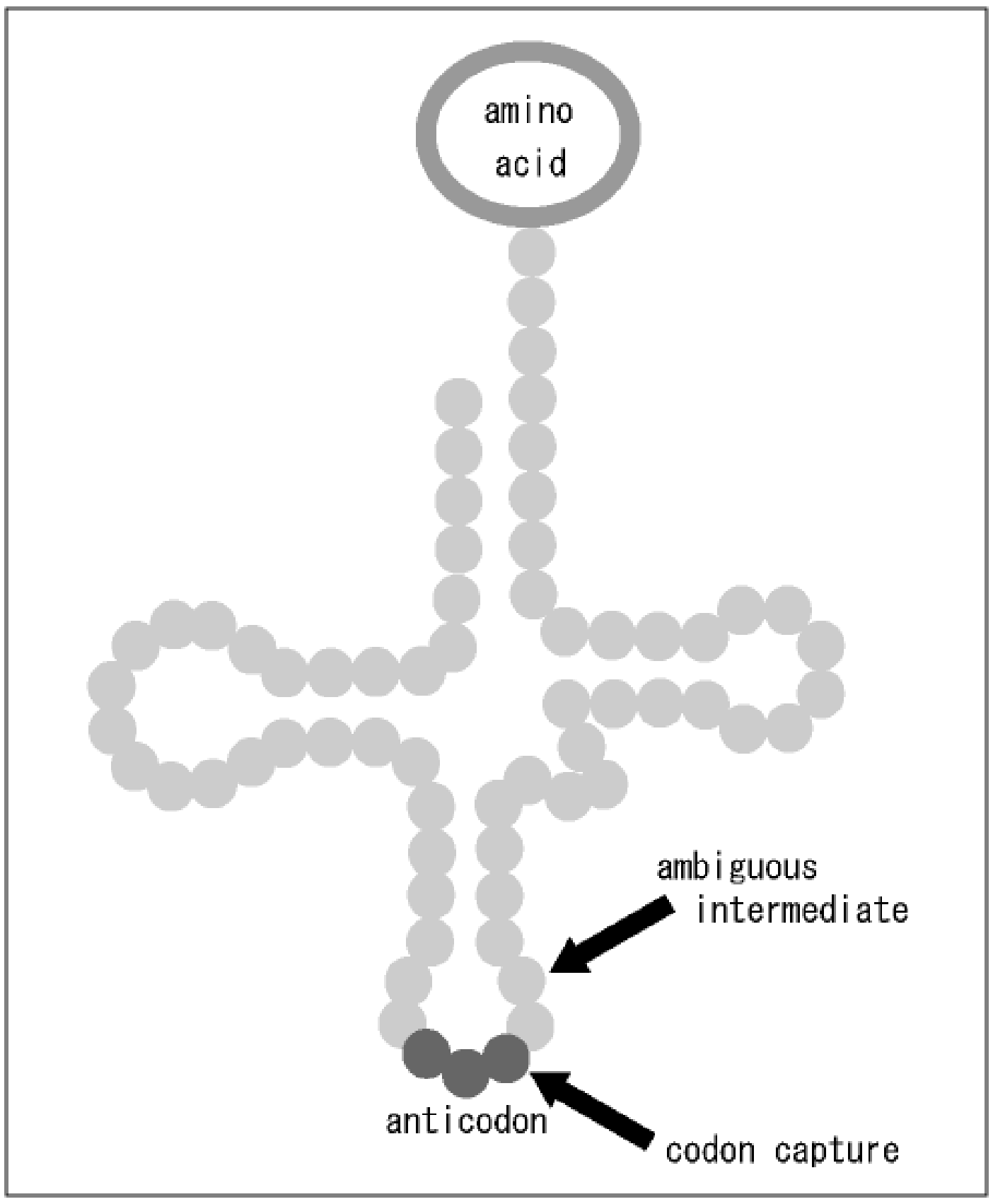

**Fig. 2** Mutations in tRNA. An alteration at the anticodon in tRNA is the driving force of the codon capture scenario, while the ambiguous intermediate scenario is driven by that at locations other than the anticodon.

Another important basis of our model is the following discrimination [7] between two scenarios. The codon capture scenario is supported by observed codon reassignments from stop to an amino acid. On the other hand, the ambiguous intermediate scenario is supported by observed codon reassignments from one amino acid to another.

On these two bases we introduce a concept of codon level and discuss its structure as a key ingredient to obtain a unified framework for the codon capture and ambiguous intermediate scenarios.

2-2 Codon Level

Here we model the affinity of tRNA for codon. The property of the target codon can be represented as a vector in a high-dimensional coordinate space. The coordinates constitute of the information of the properties of the codon such as the form and the physical quantities. For example, the charge or hydrophobicity distribution of the codon is such a physical quantity. Since the purpose of this paper is to sketch our scheme, we adopt a scalar, representing the property of the target codon, instead of a vector, for simplicity. In the following we call this scalar the codon level. This simplified model is a minimal model to discuss the mechanism of the genetic code evolution.

A similar construction [8] has been done in the study of immunity and a shape space is employed to describe the process of recognition between antigens and antibodies. As the first step the 1-dimensional shape space has been thoroughly investigated there.

On the above-mentioned two bases we can naturally assume a structure in the distribution of the codon levels. The structure consists of intermittent scatter of clusters as shown in Fig. 3. The cluster corresponds to the family box of four codons in the genetic code table. These codons in the same family are often translated to the same amino acid by wobble base-paring between codons and anticodons. Such a wobble pairing is considered to result form two factors. One is the closeness of the codon levels in the cluster. The other is the spread of the range of the codon recognition of tRNA. This range is expressed as a section in the coordinate axis along the codon level as shown in Fig. 3. If a substructure with level gap exists in the cluster, codons separated by the gap may be translated to different amino acids.

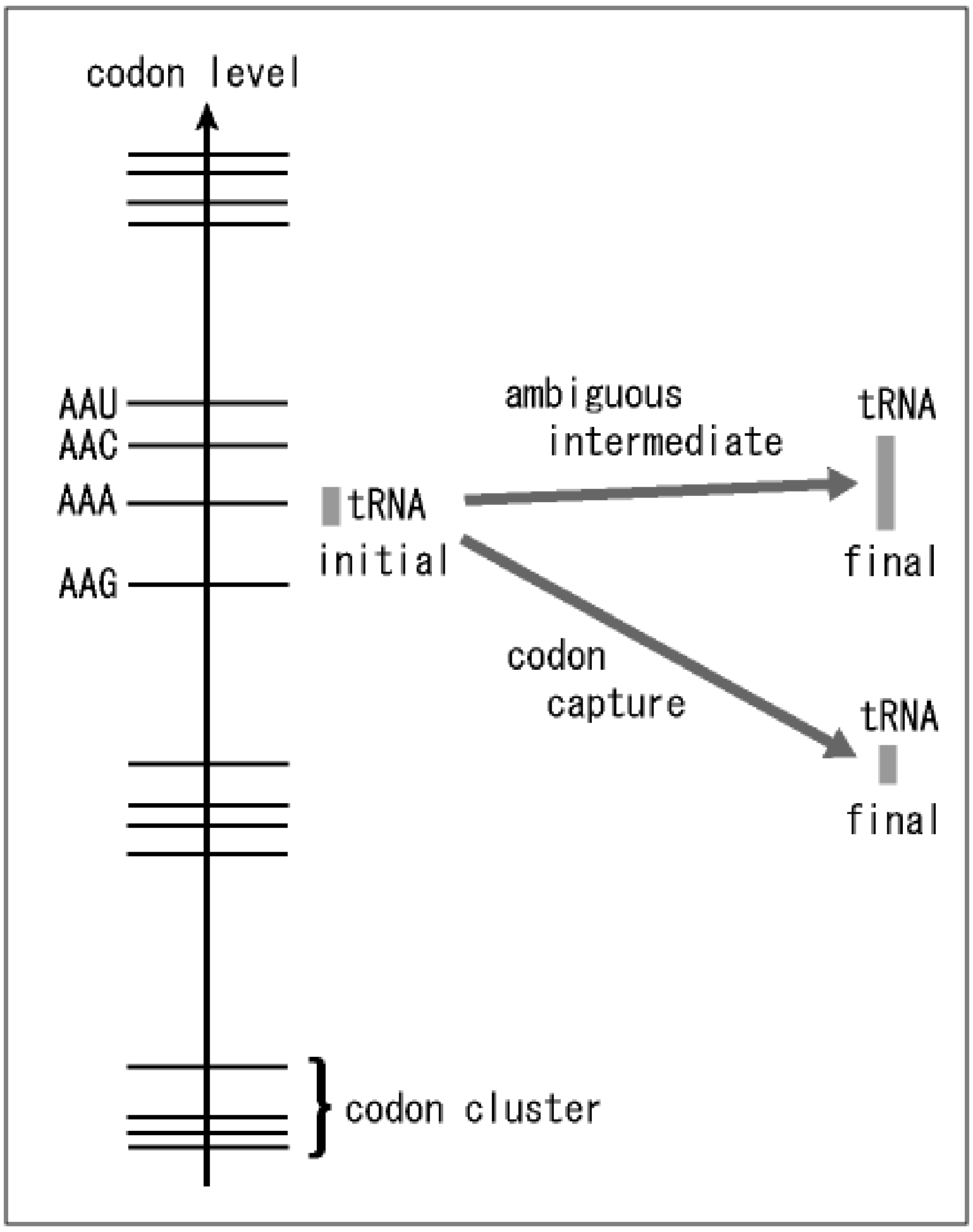

**Fig. 3** Codon level scheme. In general the levels of four codons, belonging to the same family box of the genetic code table, form a cluster and clusters belonging to different families are separated by gaps in the level scheme. The range of the codon recognition of a tRNA is expressed as a section along the codon level axis. Codon capture and ambiguous intermediate scenarios are understood as different evolutional routes. The former scenario corresponds to the genetic code evolution via inter-cluster route and the latter via intra-cluster route.

In our codon level scheme both codon capture and ambiguous intermediate scenarios are naturally understood as different evolutional route in the same coordinate space. The former scenario corresponds to the genetic code evolution via inter-cluster route and the latter via intra-cluster route. Since the change in the range of codon recognition of tRNA by an alteration at anticodon is so large to exceed the level gap between codon clusters, the inter-cluster route arises from an alteration at the anticodon in tRNA. On the other hand, the intra-cluster route arises from an alteration at locations other than the anticodon in tRNA. The change in the range of codon recognition of tRNA by this type of alteration is smaller than the level gap.

Since the needed coordinate space is smaller and the mechanism of the selection among competing tRNAs can be implemented explicitly, we simulate the ambiguous intermediate scenario in the next section.

## 3 Simulation

Here we perform some simulations on the basis of our codon level model focusing on the smallest part of the level scheme in which we can realize the ambiguous intermediate scenario. We pursue the time dependence of the tRNA distribution around two codon levels in a codon cluster.

Our algorithm for the time evolution of tRNA is as follows. Step 0: We prepare the system consisting of two homogeneous groups of tRNAs. Step 1: We introduce mutations into tRNAs. Step 2: We eliminate ill-conditioned tRNAs with too narrow or too wide range of codon recognition. Step 3: We replicate tRNAs according to the cost of the synthesis of corresponding amino acids. Step 4: We divide the system into two, mimicking cell division, when the number of tRNAs reach a critical value. We repeat the procedure from Step 1 to Step 4 sequentially.

In the following we describe the detail of the simulation.

Step 0: As the initial condition for the simulation, the 0-th generation of the system, we prepare two homogeneous groups of tRNA, tRNA-1 and tRNA-2. The initial number of tRNAs of each group is $N_0$. The range of codon recognition for each tRNA, $t_i$, is given as $C_1 - W_1 \le t_i \le C_1 + W_1$ for tRNA-1 and $C_2 - \Delta - W_2 \le t_i \le C_2 - \Delta + W_2$ for tRNA-2 where each tRNA is labeled by number $i$. The codon level for codon-1 is $C_1$ and the center of $t_i$ for tRNA-1 is chosen as $C_1$. The codon level for codon-2 is $C_2$ and the center of $t_i$ for tRNA-2 is chosen as $C_2 - \Delta$. The values of the centers, $C_1$ and $C_2 - \Delta$, are assumed to be determined by the anticodons in tRNAs and constants in the following. The constant shift $\Delta$ represents the possibility that tRNA-2 may be taken over by tRNA-1. The widths of the range, $W_1$ and $W_2$, are constants at initial stage so that the groups of tRNA-1 and tRNA-2 are homogeneous.

Step 1: We introduce random mutation into randomly selected tRNA and give new width of the range, $w_i$, of codon recognition. Such mutations are assumed to occur at locations other than the anticodons in tRNAs and change only widths of the recognition range keeping the center values unchanged. If a mutation occurs at the anticodon in tRNA, the center value is drastically changed and the recognition range is pushed out of the codon cluster now investigated. Such a mutation corresponding to the codon capture scenario is not considered here. The rate of the mutation, $\mu$, is fixed as $\mu = N / N_0$ where $N$ is the number of selected tRNAs as mutants. In order to represent the variety of nucleotides we use a weight function $f(x) = w_0 \cdot [\frac{2}{L}(x - \frac{L}{2})]^{2\alpha+1}$ shown in Fig. 4 and mutate $w_i$ into $w_i + f(x_i)$ where $x_i$ is a random number generated for each tRNA. With this mutated width $w_i$ new range of recognition, $t_i$, is given as $C_1 - w_i \le t_i \le C_1 + w_i$ for tRNA-1 and $C_2 - \Delta - w_i \le t_i \le C_2 - \Delta + w_i$ for tRNA-2.

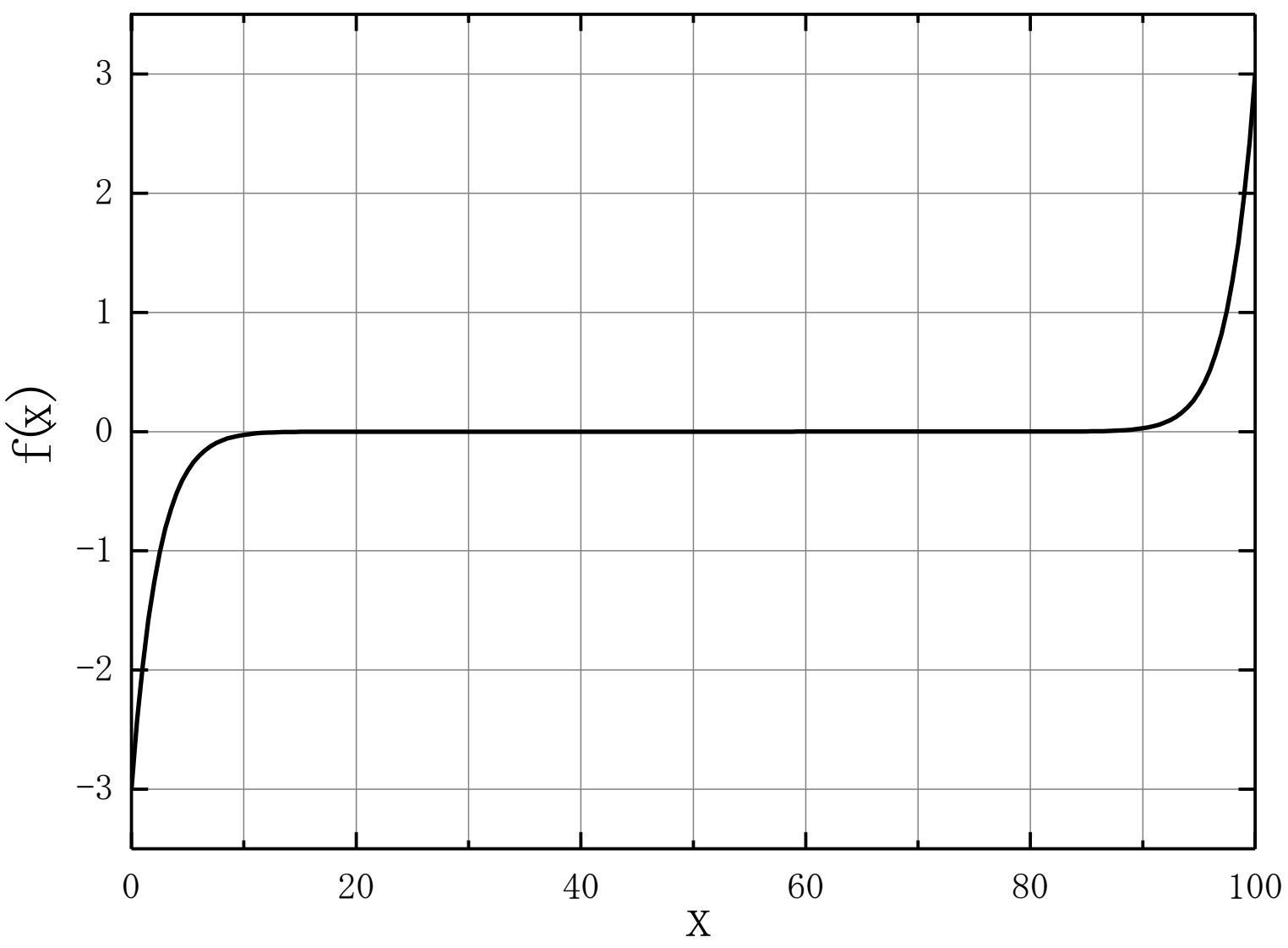

**Fig. 4** Weight function $f(x) = w_0 \cdot [\frac{2}{L}(x - \frac{L}{2})]^{2\alpha+1}$ with $L = 100$.

Step 2: We evaluate the effect of the mutation by keeping only tRNAs satisfying the condition $W_{\min} \le w_i \le W_{\max}$ which guarantees the ability of codon recognition. The lower value $W_{\min}$ represents a tolerance in order to function in some fluctuating environment not explicitly taken into account in our present model. The upper value $W_{\max}$ guarantees a singular nature of the recognition. The other tRNAs with poor recognition ability are eliminated.

Step 3: We replicate surviving tRNAs and expand the system. The rate of the replication is determined by the cost of the corresponding amino acid production. Thus we explicitly implement the selection among competing tRNAs. The amino acids transported by tRNA-1 and tRNA-2 are named aa-1 and aa-2, respectively. Since we focus on the case where tRNA-2 is taken over by tRNA-1, the rapidity, $r_1$, of the replication for aa-1 is set to be larger than that, $r_2$, for aa-2. The original tRNA for the replication is chosen randomly.

Step 4: When the total number of tRNAs reach the critical value, $4N_0$, we make new systems at the next generation by dividing the system into two, each of which consists of randomly selected $2N_0$ tRNAs, and pursue one of the new systems afterward. The system of the $g$-th generation results from $g$ times divisions.

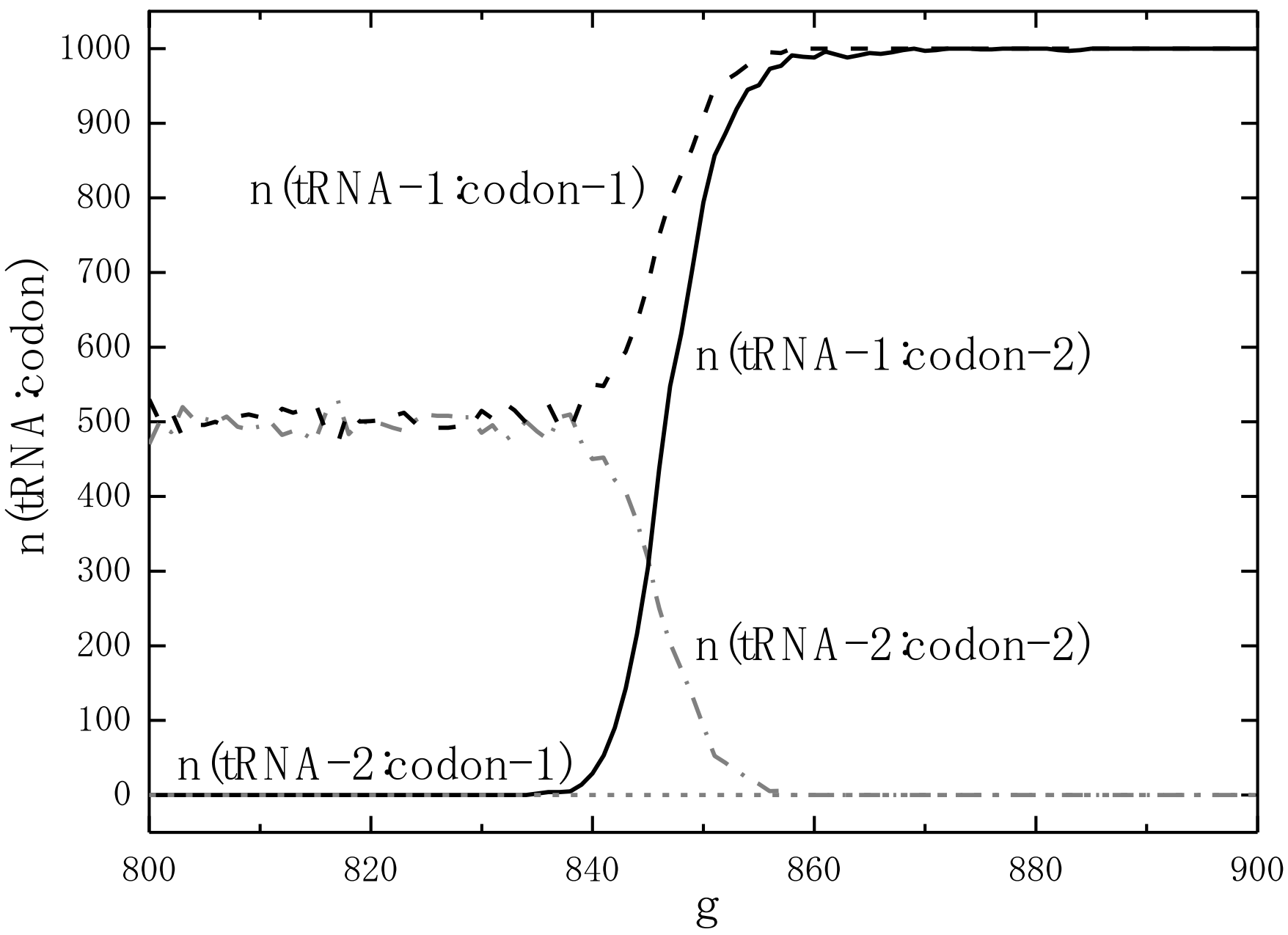

**Fig. 5** Codon assignment transition for $N_0 = 500$, $C_1 = 15$, $C_2 = 10$, $\Delta = 1$, $W_{\max} = 8$, $W_{\min} = 0.1$, $\mu = 10^{-2}$. The number of tRNA is represented, for example, as *n*(tRNA-1:codon-2) which is the number of tRNA-1 recognizing codon-2. In the initial state all codon-1 are recognized by tRNA-1 and all codon-2 by tRNA-2. In the final state all of codon-1 and codon-2 are recognized by tRNA-1, since we have chosen as $r_1 / r_2 = 1.5$. Here $g$ represents the generation defined by the number of cell division.

In Fig. 5 an example of the time dependence of the codon recognition of tRNAs is shown. The tRNA with higher cost of amino acid production, tRNA2, is taken over by the other, tRNA1. For saving the simulation time the mutation rate is set to be high as $\mu = 10^{-2}$. Once a tRNA which recognizes both codons appears the takeover rapidly progresses irrespective of the mutation rate. The rate controls the starting time of the takeover.

The period of the ambiguous translation is independent of the cost of amino acid production if the costs for two amino acids differ considerably as shown in Fig. 6. Since slight difference of the reactions leads to drastic change in the characteristic time of amino acid production, it is taken for granted that the costs differ significantly when the takeover occurs. Thus the period is about 10 generations and very short in comparison with the time scale of evolution. Such rapidness of the takeover reduces uncomfortableness of the non-unique translation of the code at intermediate state that is the weakness of the ambiguous intermediate scenario.

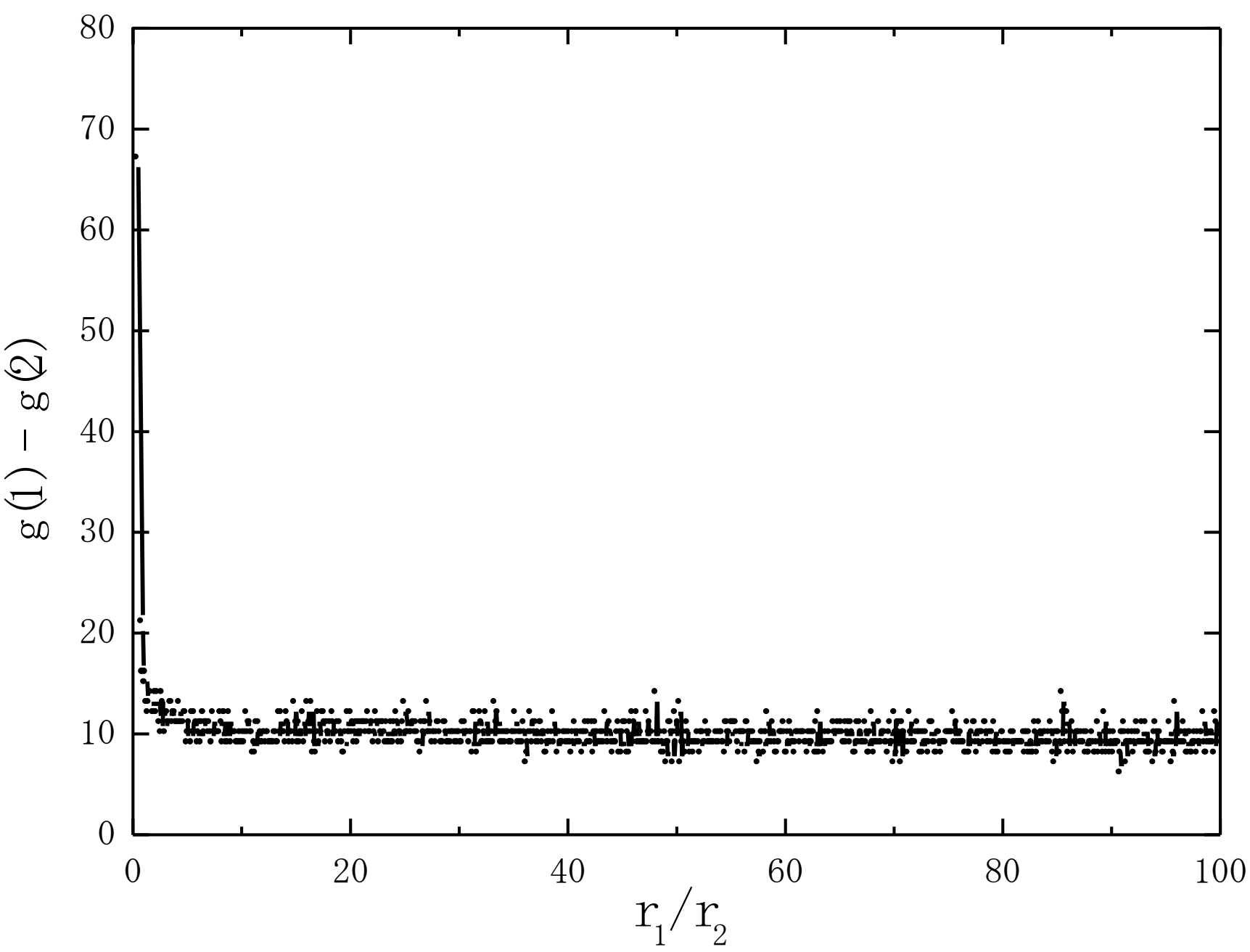

**Fig. 6** Transition time as a function of $r_1 / r_2$ for the simulation in Fig. 5. At the generation $g(2)$, *n*(tRNA-1:codon-2) first exceeds 50 and at $g(1)$, it first exceeds 950. The transition time is estimated by $g(1) - g(2)$.

## 4 Conclusion

We have proposed a physical model to discuss the mechanism of the genetic code evolution. In the framework of our simplified model, a minimal model, the codon capture and ambiguous intermediate scenarios are understood in a unified manner. On the basis of the hierarchical structure of the codon levels two scenarios are classified into two different routes of the evolutional process.

Our simulation implemented cost selection of amino acids has demonstrated a rapid transition of the code change in the case of the ambiguous intermediate scenario. This rapidness of the transition covers the drawback, the non-unique translation of the code during the transition, of the scenario. It should be noted, however, that the cost is only one of the selection pressures and we have to consider the selection from more global point of view for the environment of cells in future study.

Since the aim of our paper is to illustrate the hierarchical structure of the coordinate space of tRNA, our model is too simple and abstract. Although we should employ more realistic model in order to describe biological phenomena, such elaboration is left to future study.

Another attempt [9] to unify the codon capture and ambiguous intermediate scenarios has been reported. Such a probabilistic approach does not reach the mechanism, which is our main subject, of the genetic code evolution.

## References

[1] Osawa S (1995) Evolution of the genetic code. Oxford University Press, Oxford

[2] Knight RD, Freeland SJ, Landweber LF (2001) Rewiring the keyboard: evolvability of the genetic code. Nature Rev. Genet. **2**, 49-58

[3] Knight RD, Landweber LF, Yarus M (2001) How mitochondria redefine the code. J. Mol. Evol. **53**, 299-313

[4] Osawa S, Jukes TH (1989) Codon reassignment (codon capture) in evolution. J. Mol. Evol. **28**, 271-278

[5] Schultz DW, Yarus M (1994) Transfer RNA mutation and the malleability of the genetic code. J. Mol. Biol. **235**, 1377-1380

[6] Schultz DW, Yarus M (1996) On malleability in the genetic code. J. Mol. Evol. **42**, 597-601

[7] Swire J, Judson OP, Burt A (2005) Mitochondrial Genetic Codes Evolve to Match

Amino Acid Requirements of Proteins. J. Mol. Evol. **60**, 128-139
[8] Perelson AS, Weisbuch G (1997) Immunology for physicists. Rev. Mod. Phys. **69**, 1219-1267
[9] Sengupta S, Higgs PG (2005) A Unified Model of Codon Reassignment in Alternative Genetic Codes. Genetics **170**, 831-840